# Polar nano-clusters in nominally paraelectric ceramics demonstrating high microwave tunability for wireless communication

Hangfeng Zhang, [a, b] Henry Giddens,[a] Yajun Yue,[b] Xinzhao Xu,[b] Vicente Araullo-Peters, [c] Vladimir Koval, [d] Matteo Palma, [b] Isaac Abrahams, [b] Haixue Yan, [c, *] and Yang Hao [a, *]

[a] School of Electronic Engineering and Computer Science, Queen Mary University of London, Mile End Road London E1 4NS, UK

[b] School of Biological and Chemical Sciences, Queen Mary University of London, Mile End Road London E1 4NS, UK.

[c] School of Engineering and Materials Science, Queen Mary University of London, Mile End Road London E1 4NS, UK.

[d] Institute of Materials Research, Slovak Academy of Sciences, Watsonova 47, 04001 Kosice, Slovakia.

**Corresponding Authors**

*Haixue Yan: h.x.yan@qmul.ac.uk

*Yang Hao: y.hao@qmul.ac.uk

KEYWORDS: microwave, dielectric, tunability, polarization, ceramics

ABSTRACT: Dielectric materials, with high tunability at microwave frequencies, are key components in the design of microwave communication systems. Dense $Ba_{0.6}Sr_{0.4}TiO_3$ (BST) ceramics, with different grain sizes, were prepared in order to optimise the dielectric tunability via polar nano cluster effects. Dielectric permittivity and loss measurements were carried at both high and low frequencies and were supported by results from X-ray powder diffraction, scanning and transmission electron microscopies, Raman spectroscopy and piezoresponse force microscopy. The concentration of polar nano clusters, whose sizes are found to be in the range 20 to 50 nm, and the dielectric tunability increase with increasing grain size. A novel method for measurement of the microwave tunability in bulk dielectrics is presented. The highest tunability of 32% is achieved in ceramics with an average grain size of 10 μm. The tunability of BST ceramics with applied DC field is demonstrated in a prototype small resonant antenna.

**Introduction**

Wireless communications have become an integral feature of modern day life, as innovations such as 4[th] and 5[th] generation cellular networks and WIFI allow an increasing number of devices to be connected to the internet. With this trend set to continue, various research communities have been developing novel technologies to maximise the capacity of wireless communication systems. One area, that has been the focus of significant interest, is active and reconfigurable microwave components, such as tunable antennas [1,2] and meta-surfaces [3]. Reconfigurable antennas, for example, are able to control the frequencies at which they operate, and/or the directions in which power is radiated. In conjunction with recent advances in cognitive and software-defined radio systems, active microwave antennas can be utilized within wireless communications systems that are able to operate over wide frequency ranges, or that have the ability to track mobile targets.

One method of controlling the propagation characteristics of electromagnetic waves is through tunable materials. Among the various materials that have been studied for tunable microwave applications [4–6], ferroelectric materials, which transform to a nominally paraelectric state above their Curie point, are promising candidates, due to their high dielectric permittivity and tunability [4,7]. Although there has been much research on polarisation and domain structure in ferroelectric phases, the existence of local polar structures in nominally paraelectric phases has been rarely reported [8–15]. In some cases, despite being in a nominally paraelectric state (i.e. above the Curie temperature), high dielectric permittivity and tunability are observed, indicative of a local polar structure. One of the most intensively studied systems for tunable microwave applications is the barium strontium titanate solid solution ($Ba_xSr_{1-x}TiO_3$), which exhibits paraelectric behaviour at room temperature in compositions with $x < 0.3$ [16,17] and has shown promising tunable dielectric properties for microwave applications at room temperature [18–22]. Most of this research has focussed on dielectric measurement and properties of thin films [23,24]. This is difficult to replicate at microwave frequencies, as the longer wavelength requires larger structures, whilst it is not possible to measure free-space transmission/reflection on ceramic samples which have a diameter that is typically much lower than the wavelength. Moreover, the properties of these films are greatly influenced by the interface with the substrate, which can, for example, alter the Curie point and polarisation dramatically [12,13].

Microwave applications can be based on both film and bulk materials. Compared to thin films, bulk ceramics better represent the basic material, allowing for research into the tuning mechanism of the material itself. Measurement of microwave tunability in bulk materials is rarely reported, due to the challenge of measurement of the microwave permittivity in an applied DC field. In order to maximise the potential of tunable ceramic dielectrics in microwave devices, a better understanding of the mechanisms contributing to dielectric tunability at this frequency range is required. In this paper, grain size control in dense BST ceramics has been used to optimise microwave tunability through elastic behaviour of polar nano-clusters. Uniquely, we have been able to visualize these polar nano-clusters in a bulk ceramic through piezoresponse force microscopy (PFM). Additionally, in order to measure the microwave tunability in these bulk ceramics, a novel characterisation method has been developed utilising a modified complementary split ring resonator (CRSS) with an applied DC field. Finally, the tunability of these ceramics is demonstrated in a prototype electrically small, tunable BST-based antenna.

**Experimental**

$Ba_{0.6}Sr_{0.4}TiO_3$ (BST) powder was prepared by solid state synthesis. Stoichiometric amounts of barium carbonate (Aldrich, 99.999%), strontium carbonate (Aldrich, 99.9%) and titanium dioxide (Aldrich, 99.8%) were milled in a nylon jar using ethanol and zirconium oxide balls. Milling was performed in a planetary ball mill (QM-3SP4, Nanjing University Instrument Plant, China) at 360 rpm for 4 h, followed by drying. The dried powder was sieved through a 250 μm sieve and then calcined at 1100 °C for 4 h. After cooling, the calcined powder was subjected to a further milling, drying and sieving process. The resulting powder was mixed with a PVA binder and uniaxially pressed into pellets of 15-40 mm in diameter and ca. 1.5 mm in thickness under a pressure of 200 MPa. The pellets were sintered at temperatures ranging from 1200 to 1500 °C for 3 h with a heating/cooling rate of 3 °C min$^{-1}$.

The pellets were coated with gold and their morphology was examined by scanning electron microscopy (SEM, FEI Inspect-F Oxford). The pellets were crushed and ground into fine powder for the X-ray powder diffraction (XRD) measurements. XRD data were collected on a PANalytical X'Pert Pro diffractometer, fitted with an X'Celerator detector using Ni filtered Cu-Kα radiation (λ = 1.5418 Å). Data were collected at room temperature in flat plate θ/θ geometry over the 2θ range 5 - 120°, with a step width of 0.0334° and an effective count time of 50 s per step. The low temperature XRD measurements for the studied samples were carried out using a Rigaku Ultima IV powder diffractometer, equipped with a low-temperature chamber (Anton Paar TTK 450). These data were collected at -25 °C in the 2θ range 5 - 120°, with a step width of 0.02°. X-ray patterns were modelled by Rietveld analysis using the GSAS suite of programmes [25]. Structure refinement was carried out using a standard cubic perovskite model in space group *Pm*-3*m* [2], for the room temperature data and a tetragonally distorted perovskite structure in space group *P*4*mm* [26] for the data at low temperature. Transmission electron microscopy (TEM,

JEOL 2010, Akishima, Tokyo, Japan) was carried out on BST pellets, which were mechanically ground and polished to 200 μm thickness. They were then cut into 3 mm diameter discs using a Gatan ultrasonic cutter. The discs were then dimple ground to 20 μm thickness and finished with ion milling until electron transparency was achieved. Atomic force microscopy (AFM, Bruker Dimension Icon atomic force microscope, piezoresponse force mode) was used to characterise the local polar structures using a conductive probe (SCM-PIT-V2 probe). Raman scattering spectra were collected at room temperature on BST pellets using in a Via$^{TM}$ confocal Raman microscope with a 633 nm wavelength laser.

For dielectric measurements below 1 MHz, the top and bottom surfaces of pellets were coated with silver paste (Gwent Electronic Materials Ltd. Pontypool, U.K.) and heated at 250 °C to form a conductive layer as an electrode. A precision impedance analyser (Agilent 4294A) was used to measure the dielectric permittivity and loss over the frequency range 100 Hz to 1 MHz at room temperature. The temperature dependent dielectric properties were measured using an LCR meter (Agilent 4284A). Ferroelectric current-electric field (I-E) and polarization-electric field (P-E) loops were measured with a ferroelectric hysteresis measurement tester (NPL, UK), with voltage applied in a triangular waveform at 10 Hz.

The complex microwave dielectric properties of the BST samples were measured at a single frequency point using the dielectric perturbation technique, with samples loaded inside a resonant microwave cavity, resonant at 5.051 GHz [27]. When positioned in the centre of the cavity, the frequency shift of the TE$_{111}$ mode is directly related to the real part of the permittivity, whilst the imaginary part of the permittivity affects the Q-factor of the cavity. By comparing the measured data with numerical modelling of the loaded cavity, the complex permittivity of each sample was extracted within an error of ± 5%. The frequency at which the dielectric permittivity was measured was determined from the frequency shift of the resonant cavity mode caused by the sample. For pellets of 13.3 ± 0.1 mm diameter and 0.7 ± 0.1 mm thickness, this measurement frequency was approximately 3.8 GHz.

The measurement of the broadband dielectric properties at microwave frequencies was carried out using an open ended coaxial probe, with an outer diameter of 20 mm (Agilent/Keysight 8057E). The reflection coefficient was measured using a Network Analyser (Agilent/Keysight PNA-L N5230C) and the dielectric constant was extracted using the Keysight Material Measurement Software (N1500A materials measurement suite). The coaxial probe was calibrated by placing it in contact with deionized water as a reference. For the measurements, the inner conductor and outer ground of the coaxial probe were placed in contact with the BST samples at a number of different positions (a minimum of 8 measurements on both sides of the sample). In these measurements, only the real part of the permittivity was extracted due to the large uncertainties in the measurement of the imaginary component for solid materials.

For the tunability measurements at microwave frequencies, an in-house complementary split ring resonator (CSRR) loaded transmission line permittivity sensor was used [28]. Silver electrodes were applied to the top surface of the pellets as described above. The samples were placed on the CSRR transmission line sensor and the two-port S-parameters were measured using a Rohde and Schwarz ZNBT8 Vector Network Analyser. A DC voltage was applied using a FUG MCP 1400-1250 DC power supply in 25 V incremental steps from -850 V to +850 V. A 10 s settling time was allowed before recording the data at each step. The frequency at which the tunability was measured is related to the shift in resonant frequency when the sample is placed on the sensor (which is dependent on the relative permittivity of the individual samples, and their thickness) and for all results reported here, fell between 1 – 3 GHz, although frequency tunability was observed at higher order resonant modes at frequencies up to 8 GHz.

An electrically small resonating antenna, with reconfigurable resonant frequency based on the application of DC bias field, was designed incorporating a BST ceramic pellet. The antenna structure is similar to that reported by Wang *et al*.[29] and consists of a 50 Ω microstrip transmission line on the bottom side of a standard printed circuit board (PCB) dielectric substrate (FR-4, $\varepsilon_r$ = 4.3, tan δ = 0.02, thickness = 0.8 mm), with a width $f_w$ = 1.5 mm. On the top side of the FR-4 substrate, a small slot was etched into the continuous copper ground plane (35 µm thick copper cladding was etched away using conventional PCB techniques). A 0.55 mm thick BST ceramic disc was positioned on top of the slotted aperture, with a metallic patch on top of this. The dimensions of both the slot and patch affect the frequency at which the antenna resonates and must be similar. The slotted aperture had a width $s_w$ = 0.9 mm and length $s_L$ = 9 mm, and the patch had a width $p_w$ = 4.5 mm and length $p_L$ = 10 mm. On the non-radiating edge of the patch, a bias line was attached which was connected to the DC voltage source. The resonant patch was also etched from a 0.8 mm FR-4 superstrate. Using plastic screws, the antenna was fixed together, with the BST ceramic sandwiched between the superstrate and substrate layers. A 3D printed support layer (from material VeroClearPlus, $\varepsilon_r$ = 2.7, tan δ = 0.0026) with equal thickness (to within 0.01 mm) to the BST ceramic was used in order to ensure the BST disc was centralized around the slot and patch resonators. The input response of the antenna was measured using a Keysight 8057E network analyser with a DC voltage applied to the BST bias lines. A wideband calibration of the connecting cables was performed at the port of the antenna. The radiation patterns of the antenna were measured by recording the transmission response from a feed antenna as the antenna under test was rotated through 360° in an anechoic environment. For these measurements, the test antenna was mounted above a metallic reflector, with dimensions of 20 × 20 cm, to shield it from the mounting equipment. The gain was measured by comparing the transmission response to that of a standard gain horn antenna (ETS-Lindgren's 3115 Model).

**Results and Discussion**

Scanning electron microscopy (SEM) images of the prepared BST ceramics reveal average grain sizes ranging from *ca.* 0.8 to 25 μm depending on sintering temperature (Supporting information, Fig. S1 and Table 1), with higher sintering temperatures yielding larger grains. X-ray diffraction profiles of all studied samples are well fitted with a standard cubic perovskite structure in space group *Pm*-3*m* (Fig. S2). Pure barium titanate (BT) exhibits a tetragonal structure with a longer *c*-axis length, due to the off-centre displacement of $Ti^{4+}$ ions within the titanate octahedra, which is related to long-range ferroelectric (FE) ordering. In the present system, strontium substitution interrupts long-range FE ordering. Therefore, based on the known tunability of BST ceramics in the nominally paraelectric state, it is predicted that these materials contain polar clusters embedded within a non-polar cubic matrix. A general increase in unit cell volume with sintering temperature is observed (Fig. S3), consistent with increasing Ti reduction ($Ti^{4+} \rightarrow Ti^{3+}$, with respective ionic radii of 0.605 Å and 0.670 Å for the ions in six-coordinate geometry [30]) with increasing temperature and simultaneous creation of oxide ion vacancies for charge balance [31].

**Table 1**. Sintering temperatures, average grain size, Curie temperature and fitted Curie-Weiss parameters for BST ceramics.

| Composition | Sintering Temperature | Grain size (μm) | Relative density | $T_c$ (°C) | $T_0$ (°C) | $C$ (×10$^5$) |
|---|---|---|---|---|---|---|
| $Ba_{0.6}Sr_{0.4}TiO_3$ | 1200 °C | 0.8 ± 0.1 | 94.8% | 3.1 ± 0.5 | -8.3 | 0.95 |
| $Ba_{0.6}Sr_{0.4}TiO_3$ | 1300 °C | 1.2 ± 0.2 | 97.2% | 5.8 ± 0.5 | -5.6 | 0.93 |
| $Ba_{0.6}Sr_{0.4}TiO_3$ | 1400 °C | 10 ± 2 | 97.0% | 7.5 ± 0.5 | -7.6 | 1.00 |
| $Ba_{0.6}Sr_{0.4}TiO_3$ | 1500 °C | 25 ± 5 | 97.9% | 3.6 ± 0.5 | -26.8 | 1.49 |

The frequency dependencies of dielectric permittivity ($\varepsilon'$) and loss tangent (tan δ) at room temperature, over the frequency range 100 Hz to 1 MHz for the studied samples are shown in Fig. S4. The permittivity values are consistent with data in the literature [32]. BST ceramics with an average grain size of 25 μm exhibit the highest dielectric permittivity with strong frequency dependence and high loss tangent, which are likely related to the high concentration of oxygen vacancies caused by over sintering at 1500 °C. Samples sintered at lower temperatures show less frequency dependence of dielectric permittivity and low dielectric loss (below 0.02). The dielectric permittivity of samples sintered at temperatures below 1500 °C increased with increasing grain size, due to the decreasing grain boundary density as seen in SEM images (Fig. S1). The grain boundaries have been found to be non-ferroelectric with low dielectric permittivity [33].

Fig. 1a and Fig. S5 show the thermal dependencies of dielectric permittivity and loss at selected frequencies for the studied BST ceramics. All samples show three frequency independent dielectric anomalies corresponding to phase transitions on cooling, viz.: from cubic to tetragonal at *ca*. 0 °C, tetragonal to orthorhombic at *ca*. -65 °C and orthorhombic to rhombohedral at *ca*. -100 °C. High dielectric loss in the FE phase is mainly due to the domain walls. On heating, BST undergoes the reverse phase transitions to higher symmetry crystal structures, which lower the ferroelectric distortion and reduce the density of polar clusters. BST ceramics with larger grain size exhibit sharper dielectric anomalies. The $T_c$ values increase with increasing grain size up to 10 μm, suggesting increasing density of polar clusters. The increasing polar cluster density is attributed to not only a smaller temperature gap between the Curie point and room temperature, but also to a lower density of grain boundaries which act as non-polar defect structures. The dielectric permittivity at 10 kHz was analysed by the Curie-Weiss law, which can be expressed as:

$$\frac{1}{\varepsilon\prime} = \frac{(T-T_0)}{C} \quad (1)$$

where $T_0$ is the Curie-Weiss temperature and $C$ is the Curie-Weiss constant. The fitting results are presented in Fig. 1b, with details summarised in Table 1. For samples with grain sizes ≤ 10 μm, the Curie-Weiss temperatures were between -5 to -10 °C. The sample with 25 μm grain size gave an unusually low $T_0$ value of -26.8 °C, which may be associated with defects caused by over sintering [31].

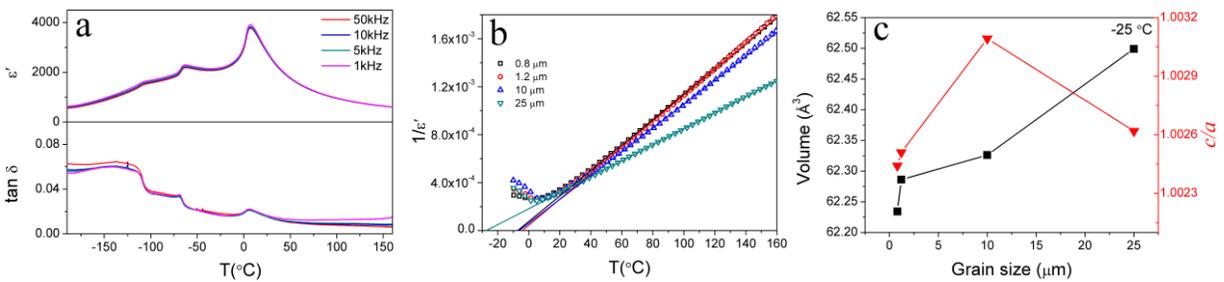

**Fig. 1.** (a) Thermal variation of the dielectric permittivity and loss tangent at selected frequencies for BST ceramics with a grain size of 10 μm; (b) Curie Weiss fitting of the dielectric behaviour of the BST samples (solid lines); (c) variation of tetragonal unit cell volume (error bars are smaller than the symbols used) and *c/a* ratio with grain size in BST ceramics.

At -25 °C XRD patterns were well fitted using a tetragonal model in space group *P4mm* (Fig. S6). As seen at room temperature, the unit cell volume increases with increasing sintering temperature (Fig. 1c). It is interesting to note that in the BST samples at -25 °C, the *c/a* ratio of the tetragonal phase increased with increasing grain size up to 10 μm, but then decreased for the 25 μm grain sized sample. It is well known that spontaneous polarization in the tetragonal structure is associated with elongation of the *c*-axis. Thus, the increasing *c/a* ratio indicates increasing levels of distortion and results in increasing spontaneous polarization. The sudden decrease in this distortion for the 25 μm grain sized sample indicates a lowering of spontaneous polarization and is likely caused by the increased defect concentration due to over sintering. Increasing distortion leads to a higher Curie point $T_c$ and this is clearly evident in Table 1, where $T_c$, increases to a maximum for the 10 μm grain sized sample.

Fig. 2 shows current-electric field (I-E) and polarization-electric-field (P-E) loops measured under an applied electric field of 10 kV mm$^{-1}$ at 10 Hz. For samples with grain sizes ≤ 10 μm, narrow P-E hysteresis loops are observed, indicative of relaxor-like behaviour, with saturation polarization increasing with increasing grain size, suggesting increasing levels of distortion within the polar clusters. This is consistent with the variation in the degree of crystallographic distortion seen in the low temperature XRD results for these samples. I-E loops for samples with grain sizes ≤ 10 μm are similar to each other, with two current peaks near zero field, associated with polarization rotation of the polar clusters [34,35]. The area within the P-E loop represents the dielectric loss at 10 Hz in the materials. The sample with 0.8 μm grain size exhibits a relatively wide P-E loop with large remnant polarization and coercive field and a slight tilting of the I-E loop, which indicates additional energy loss due to electronic conductivity[34]. The 25 μm grain sized sample became electrically conductive at 1 kV mm$^{-1}$, as a result of the high defect concentration.

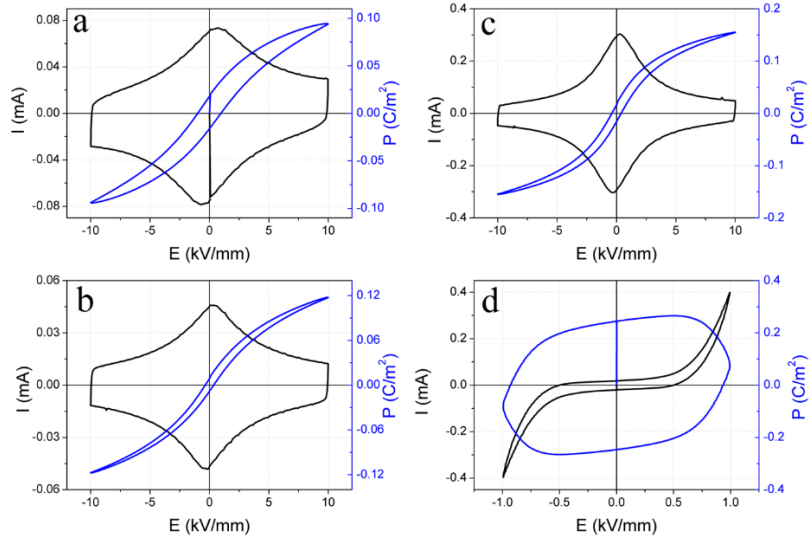

**Fig. 2**. P-E/I-E loops of BST ceramic samples measured at 10 Hz at room temperature, with grain sizes of (a) 0.8 μm, (b) 1.2 μm, (c) 10 μm and (d) 25 μm.

Fig. 3a shows the temperature dependence of dielectric permittivity of the 10 μm grain sized sample with and without an applied bias field of 1 kV mm$^{-1}$ at 1 MHz. At temperatures below *ca.* 60 °C, the dielectric permittivity of the sample with bias field is lower than that without bias field. Under DC field, the activity of the dipoles in BST is restrained and results in a decrease in dielectric permittivity. At temperatures below $T_c$, the lowering in dielectric permittivity is mainly caused by decreasing domain wall density in the FE phase induced by DC poling [35]. At temperatures above $T_c$, the difference in permittivity is dominated by the contribution of the polar clusters. Further increasing the temperature results in smaller polar clusters and their contribution becomes negligible at *ca.* 60 °C, above which temperature there is no significant change in dielectric permittivity with applied field. Transmission electron microscopy (TEM) images of the 10 μm grain sized sample (Fig. 3b) show no clear evidence for domain structure or polar clusters at room temperature. However, TEM diffraction patterns along the [120] zone axis, taken from different regions of the BST sample, show clean diffraction spots in one region (Fig. 3c), while diffuse electron scattering is observed in another region (Fig. 3d), the latter consistent with the existence of polar nano-clusters, as observed in tin doped BT systems [36]. More direct evidence of local polar nano-clusters comes from examination of the ceramics under piezoresponse force microscopy (PFM). Fig. S7 shows PFM images of a 10 μm grain sized ceramic, with the polar nano-clusters evident in the corresponding PFM phase image (Fig 3e). The polar nano-clusters appear as white or black spots in Fig. 3e and are distributed in the non-polar matrix (brown background). The size of the nano-clusters appears to be in the range of 20-50 nm. Yan *et al.* used photon correlation spectroscopy to observe the dynamics of polar nano-clusters in single crystals of BT above $T_c$ and drew the conclusion that polar nano-clusters made a major contribution to the dielectric permittivity [37]. Polar nano-clusters in BST are thought to exhibit reversible polarization with short relaxation times, which remain active at microwave frequencies and are a major contributor to the high tunability and low loss. Indeed, polar nano-clusters in single crystals of the nominally paraelectric phase of BT, observed using soft X-ray techniques, have been shown to have relaxation times of less than 100 ps [38,39].

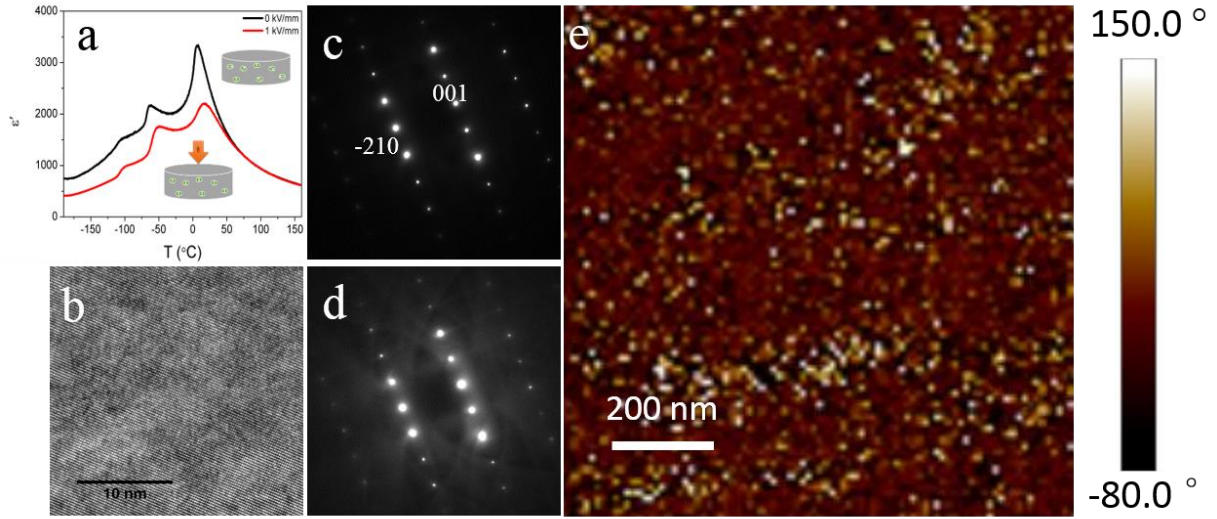

**Fig. 3.** (a) Temperature dependence of dielectric permittivity with and without bias field for a BST ceramic sample of 10 μm grain size. (b) TEM image and SAED diffraction patterns from different regions (c and d) of BST along the [120] zone axis; (e) Enlarged PFM phase image of BST sample.

Further evidence for the presence of polar nano-clusters in BST is found in the Raman spectra of BST ceramics at room temperature (Fig. 4). The spectra are all similar to each other and agree well with that of pure BT [40,41] and similar BST compositions [42]. Weak Raman peaks occurring at *ca*. 300 cm$^{-1}$ and 750 cm$^{-1}$ are associated with local tetragonal ordering in the BST ceramics, associated with the distribution of polar nano-clusters within the cubic BST matrix. In all samples, these peaks are clearly observed.

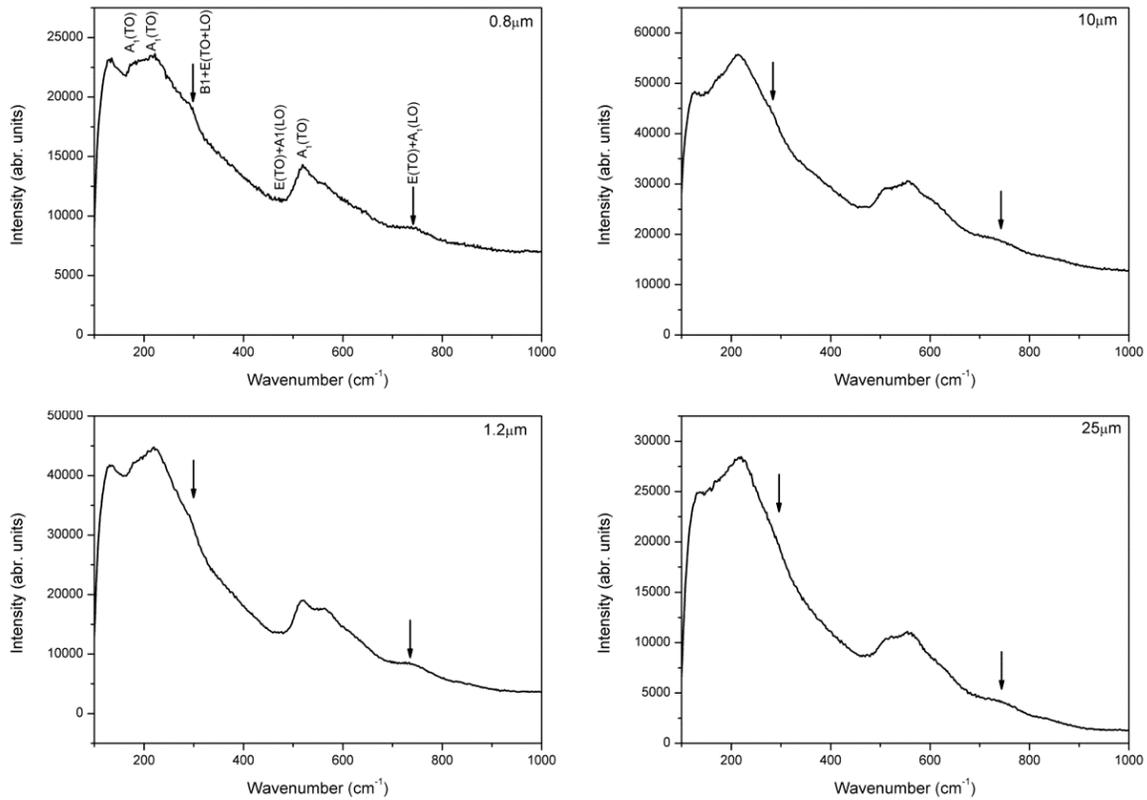

**Fig. 4.** Raman spectra of BST ceramics measured at room temperature. Raman peaks associated with the tetragonal ordering are indicated by the arrows.

The frequency dependence of dielectric permittivity for BST ceramics under DC bias field is shown in Fig. S8. When compared to the measurement at zero field, all compositions show a decrease in dielectric permittivity with applied field across the measured frequency range. As discussed above, application of a bias field, reduces the activity of the polar nano-clusters, resulting in a decrease of their contribution to dielectric permittivity. It is worth noting that several peaks appeared at high frequency (~900 kHz) and became more pronounced with increasing electric field. These peaks are possibly associated with the piezoelectric effect induced by an applied dc field [43]. The tunability ($\eta$) of a material can be expressed as:

$$\eta = \frac{\varepsilon_{0V} - \varepsilon_E}{\varepsilon_{0V}} \tag{2}$$

where $\varepsilon_{0V}$ and $\varepsilon_E$ are the permittivity at zero and applied field, respectively. As seen in Fig. S9d, samples with large grain size show slightly higher tunability with DC bias field. While there has been much work on tunability measurements at radio frequencies, there has been little in the GHz range. This is mainly due to the lack of suitable methods. The CSRR sensor used in the present study, has previously been used to determine permittivities of low permittivity materials at microwave frequencies [28]. Here, we have utilised this type of sensor for measurement of tunability, through changes in the resonant frequency under DC bias. The resonant frequency is very sensitive to changes in the permittivity and therefore we can determine tunability as:

$$\eta = \frac{f_{r(0V)}^{-2} - f_{r(E)}^{-2}}{f_{r(0V)}^{-2} - f_{r(\text{unloaded})}^{-2}} \tag{3}$$

where $f_{r(0V)}$, $f_{r(E)}$ and $f_{r(\text{unloaded})}$ are the resonant frequencies of the sensor under 0V DC bias, under applied DC bias and when the sample is removed, respectively.

Table S1 displays the microwave complex permittivity properties of the BST samples with up to 10 μm grain size, measured using the resonant cavity perturbation technique. The measurement frequencies showed dependency on size, volume and permittivity of each sample, and were recorded at approximately 3.8 GHz for all samples. The relative permittivity was in the range 130-165, with a slightly increasing trend with increasing grain size. In each case, the dielectric loss was below 0.025 for the samples with up to 10 um grain size. The losses decrease with increasing grain size, which might be attributed to the lower concentration of grain boundaries which act as defects in the material. The permittivity spectrum from dielectric probe reflection measurements was frequency independent across the 2-3 GHz range for all the studied BST ceramics (Fig. S9).

Fig. 5a shows the variation of room temperature microwave dielectric tunability with electric field for BST samples with grain sizes up to 10 μm. The tunability for the 0.8 um grain sized BST ceramic is only 3% at 1 kV mm$^{-1}$, while 1.2 μm and 10 μm grain sized samples show tunabilities of 12% and 27%, respectively. The significant variation in microwave tunability is attributed to the differences in relaxation times of the polar nano-clusters in the different grain size BST ceramics. Relaxation times will depend on the size of the polar clusters, with larger clusters exhibiting longer relaxation times. Dipoles with longer relaxation times only contribute to the dielectric permittivity at low frequencies, while dipoles with short relaxation times contribute to the permittivity at both low and high frequencies. As seen in the low and high frequency permittivity plots (Figs. S4 and S9, respectively), the permittivity values decrease from ca. 2000 at low frequencies to less than 160 at high frequencies, indicating the presence of dipoles with both short and long relaxation times. Under the same DC bias field, tunability increases with increasing grain size, reaching a maximum for the 10 μm grain sized sample with a value of 32% at 1.33 kV mm$^{-1}$. Thus, it can be predicted that the highest tunability would be achieved in single crystals, if these could be prepared with minimal defect concentration. The high tunability of the large grain size ceramics is due to the high concentration of polar nano-clusters with short relaxation times, which have a more profound impact on the dielectric tunability at microwave frequencies than dipoles with longer relaxation times. Whilst we have not measured the loss under electric field in this frequency range, it was noted that there was no significant change in the bandwidth (Q-factor) of the resonant frequency of the loaded CSRR sensor when DC-field was applied across the samples, indicating a negligible effect on the dielectric loss under electric field (Fig. 5b).

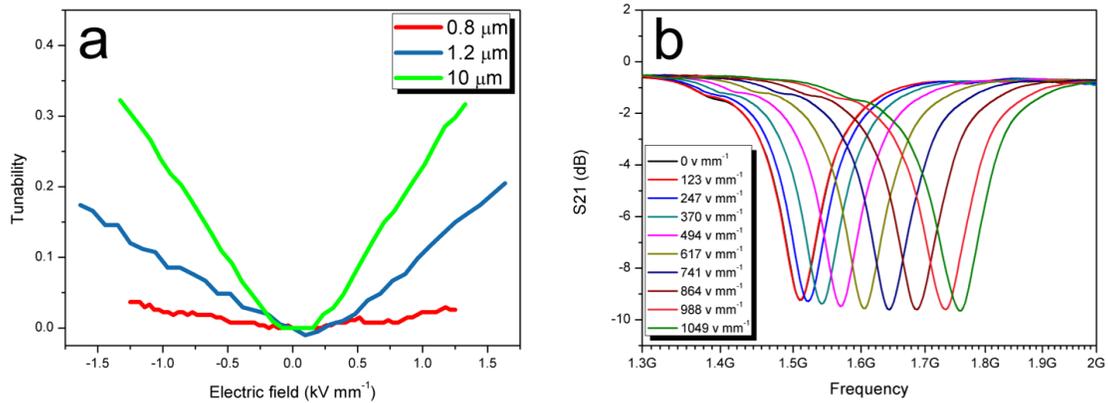

**Fig. 5.** (a) Dielectric tunability at 20 °C of BST ceramics with grain sizes ≤ 10 μm in the GHz frequency range; (b) magnitude of the transmission response of the resonant tunability sensor for a 10 μm grain sized BST ceramic under DC bias.

BST ceramics with 10 μm grain size were used as the tuning component in an electrically small, frequency reconfigurable patch antenna (Fig. 6a). Under 0 V bias, the antenna was resonant at 2.208 GHz, which increased to 2.217 GHz when the DC voltage was 400 V, and finally to 2.245 GHz when a voltage of 800 V was applied (Fig. 6b). The resonant frequency of the antenna was dependent on the bias voltage across the BST substrate and could allow for the antenna to be electrically controllable and frequency selective. Fig. 6c shows the azimuth and elevation plane patterns, normalized to the maximum point, which were stable versus frequency, as the electrical size of the antenna was always well below the wavelength. The antenna radiated with a maximum gain on -13dB due to its small electrical size of λ/13 at 0 V bias.

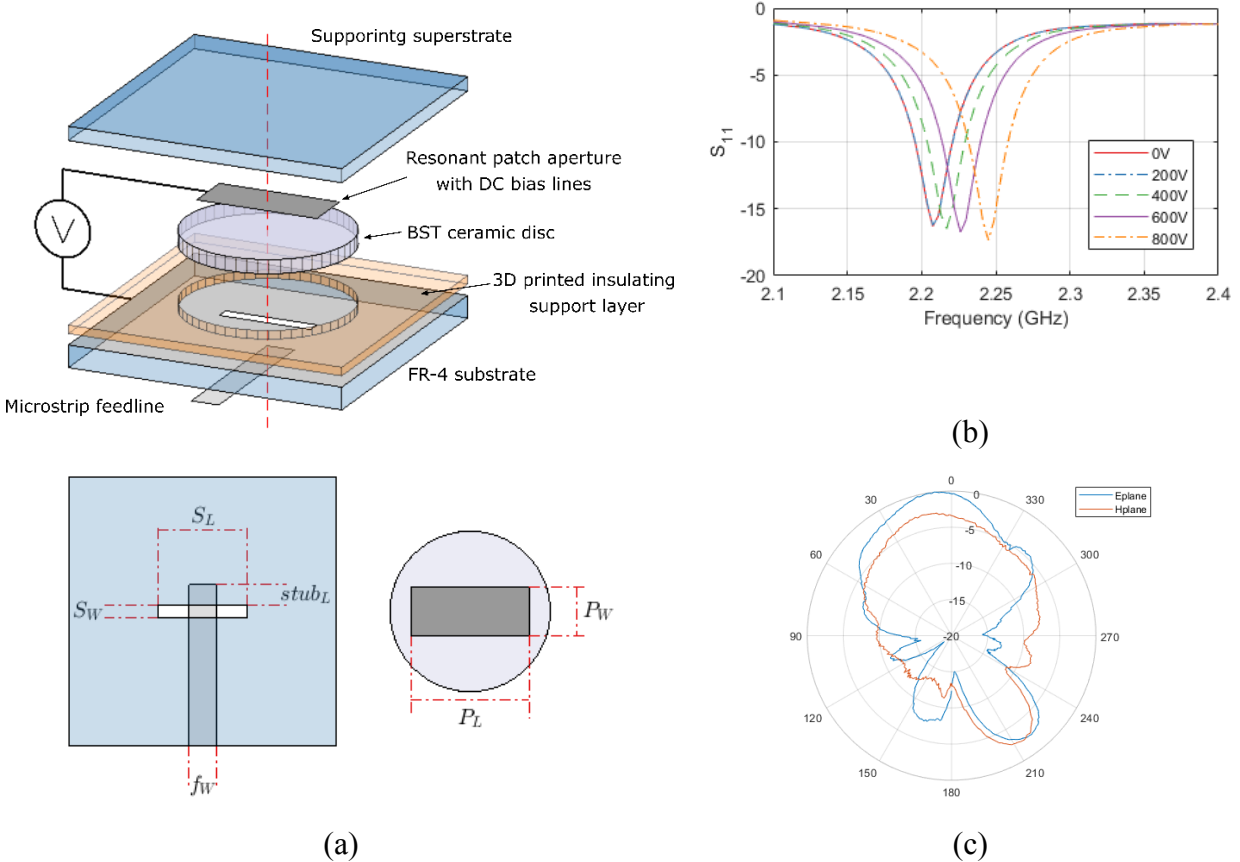

**Fig. 6.** (a) Perspective and top views of the electrically small, frequency tunable, BST based antenna; (b) measured tuning of the antenna resonant frequency with applied DC bias and (c) radiation patterns in E and H planes.

**Conclusions**

Single phase cubic perovskite BST ceramics with different grain sizes ranging from 0.8 to 25 μm have been successfully prepared through solid state methods. Samples sintered at lower temperatures (1200-1400 °C) had small grains (≤ 10 μm) and exhibited low dielectric loss and high dielectric permittivity, with tunability increasing with increasing grain size. This can be explained by the existence of polar nano-clusters in the BST ceramics, which increase in concentration with increasing grain size. These polar nano-clusters are found to have dimensions in the range 20-50 nm, with short relaxation times, and unlike other dipoles in the system are active at microwave frequencies. Thus, grain size control offers a way of tailoring tunability in these ceramics.

Using a novel method involving a modified CSRR sensor with applied DC electric field, the dielectric tunability at microwave frequencies has been measured for BST ceramics, with a sample of 10 μm grain size exhibiting the highest tunability of 32% at 1.33 kV mm$^{-1}$. Using this material an electrically small resonant antenna has been constructed and tested, demonstrating electrical control of frequency selectivity.

**Declaration of Competing Interest**

The authors declare that they have no known competing financial interests or personal relationships that could have appeared to influence the work reported in this paper.

**Acknowledgements**

This work was funded through support of DSTL under the Frequency Agile Antennas for Software Defined Radio project, by the EPSRC Animate grant (EP/R035393/1) and the Grant Agency of the Slovak Academy of Sciences (Grant No 2/0038/20). The authors would like to thank Gary Pettitt, Ian Youngs and Salman Hussain for their guidance.

**Supporting Information**.

The SI includes SEM images, XRD patterns and dielectric property data.

# Supporting Information

# Polar nano-clusters in nominally paraelectric ceramics demonstrating high microwave tunability for wireless communication

Hangfeng Zhang, [a, b] Henry Giddens,[a] Yajun Yue,[b] Xinzhao Xu,[b] Vicente Araullo-Peters,[c] Vladimir Koval,[d] Matteo Palma,[b] Isaac Abrahams,[b] Haixue Yan,[c,*] and Yang Hao [a,*]

[a] School of Electronic Engineering and Computer Science, Queen Mary University of London, Mile End Road London E1 4NS, UK

[b] School of Biological and Chemical Sciences, Queen Mary University of London, Mile End Road London E1 4NS, UK.

[c] School of Engineering and Materials Science, Queen Mary University of London, Mile End Road London E1 4NS, UK.

[d] Institute of Materials Research, Slovak Academy of Sciences, Watsonova 47, 04001 Kosice, Slovakia.

*Corresponding Authors

Haixue Yan: h.x.yan@qmul.ac.uk

Yang Hao: y.hao@qmul.ac.uk

Table S1. Dielectric properties and tunability at microwave frequency of the studied BST ceramics.

| Composition (grain size) | Permittivity | Loss | Frequency (GHz) | Tunability (1kV/mm) |
|---|---|---|---|---|
| BST (0.8 µm) | 131 ± 7 | 0.024 ± 0.004 | 3.835 | 3 % |
| BST (1.2 µm) | 141 ± 7 | 0.022 ± 0.004 | 3.865 | 12% |
| BST (10 µm) | 165 ± 8 | 0.014 ± 0.004 | 3.78 | 27% |
| BST (25 µm) | | | | N/A |

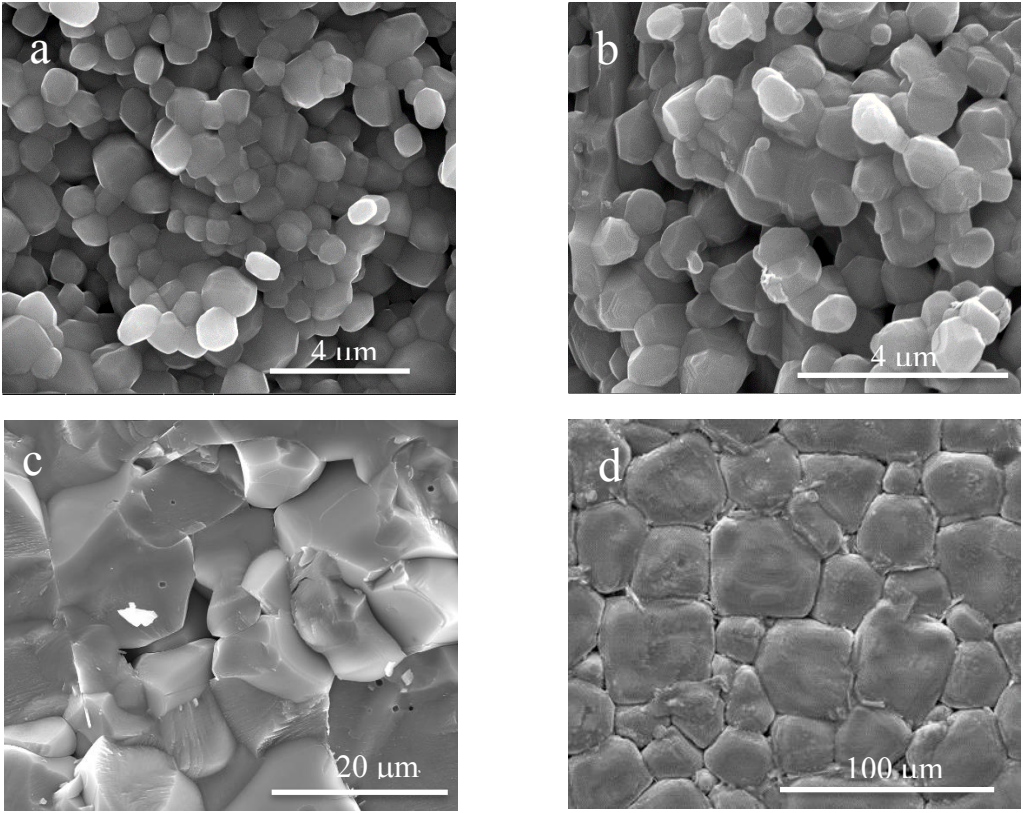

Fig. S1. SEM images of BST ceramics sintered at different temperatures; (a) 1200 °C, (b) 1300 °C, (c) 1400 °C and (d) 1500 °C.

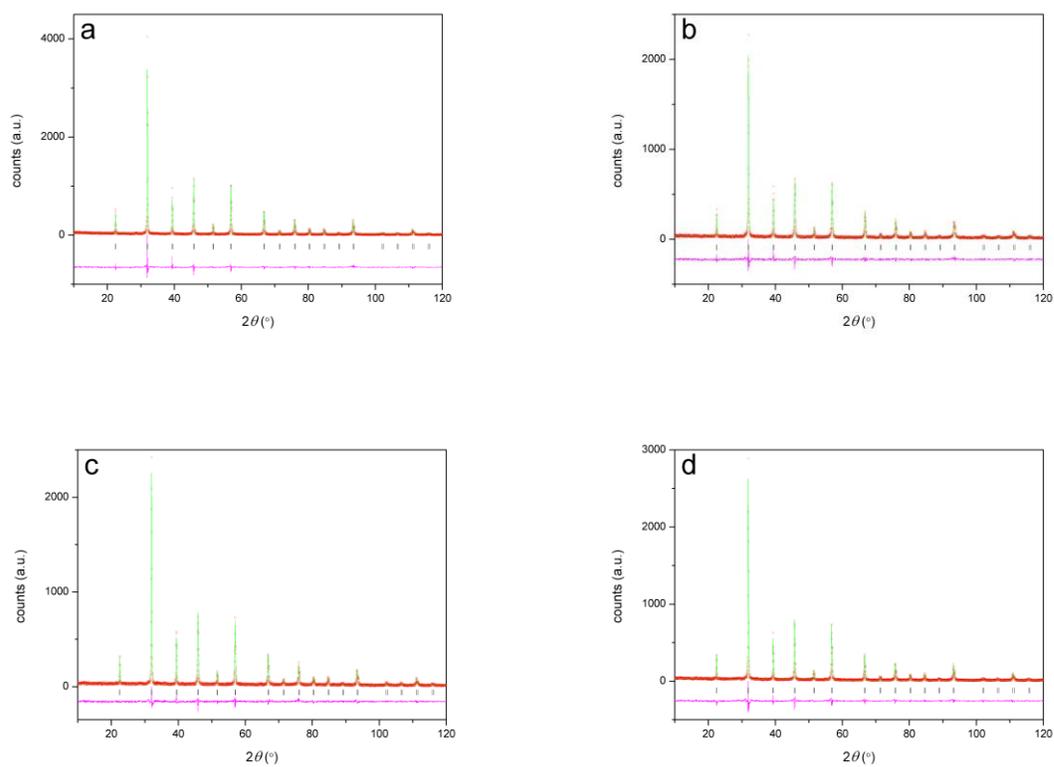

Fig. S2. Fitted X-ray powder diffraction profiles for BST ceramic samples with grain sizes of (a) 0.8 μm, (b) 1.2 μm, (c) 10 μm and (d) 25 μm. Observed (red crosses), calculated (green solid line) and difference (purple solid line) profiles are shown along with the Bragg peak positions (black markers).

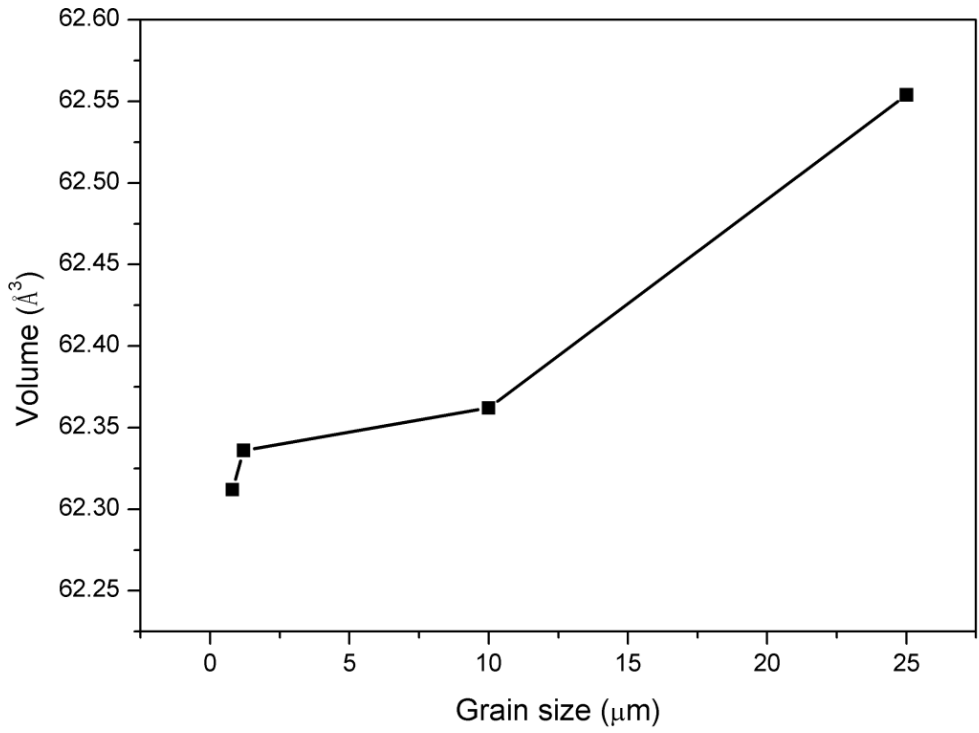

Fig. S3. Variation of cubic unit cell volume of BST ceramics with grain size.

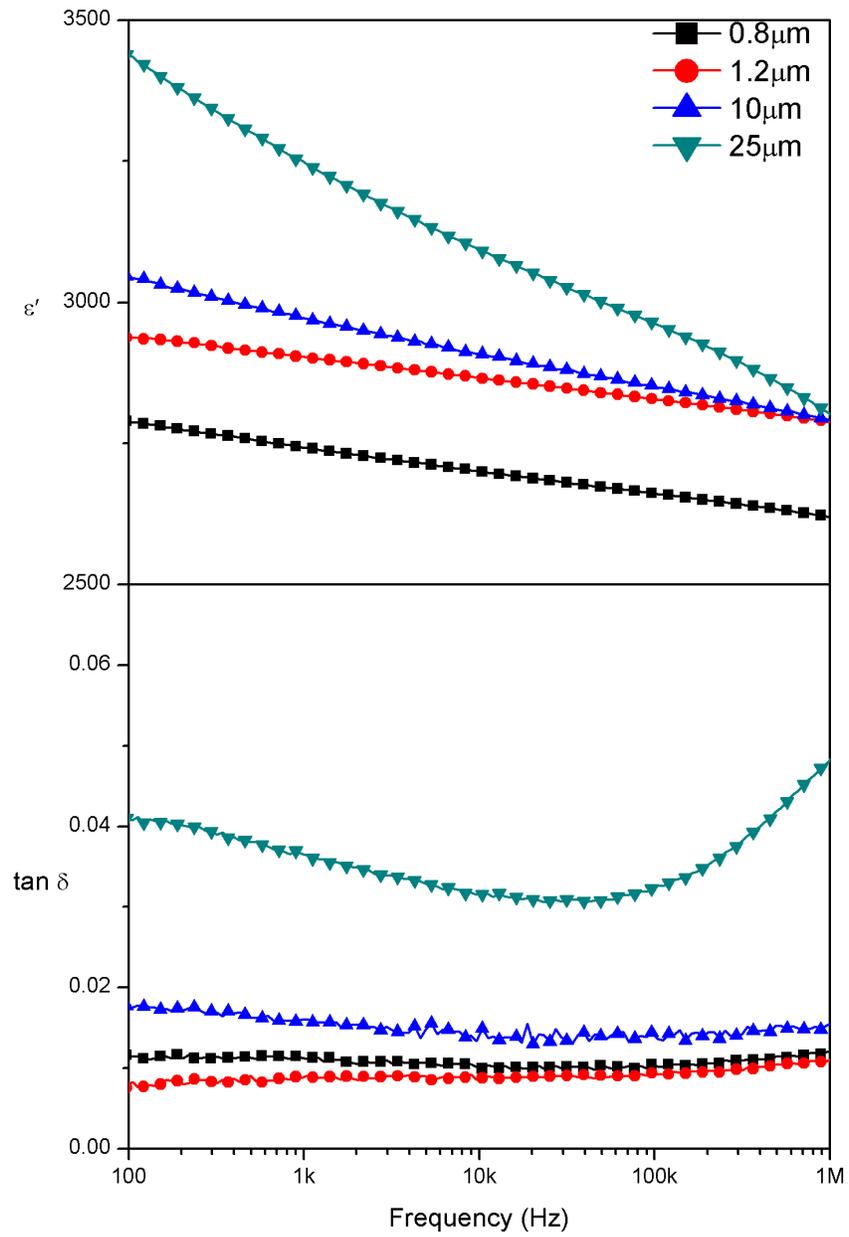

Fig. S4. Frequency dependence of dielectric permittivity and loss tangent of BST samples.

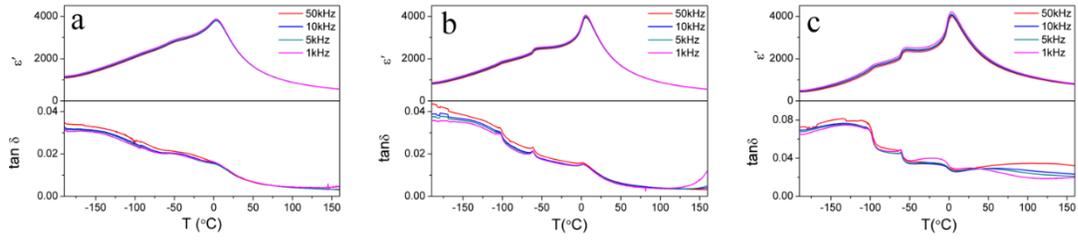

Fig. S5. Thermal variation of relative dielectric permittivity and loss tangent at selected frequencies for BST ceramics with grain sizes of (a) 0.8 μm, (b) 1.2 μm, and (c) 10 μm.

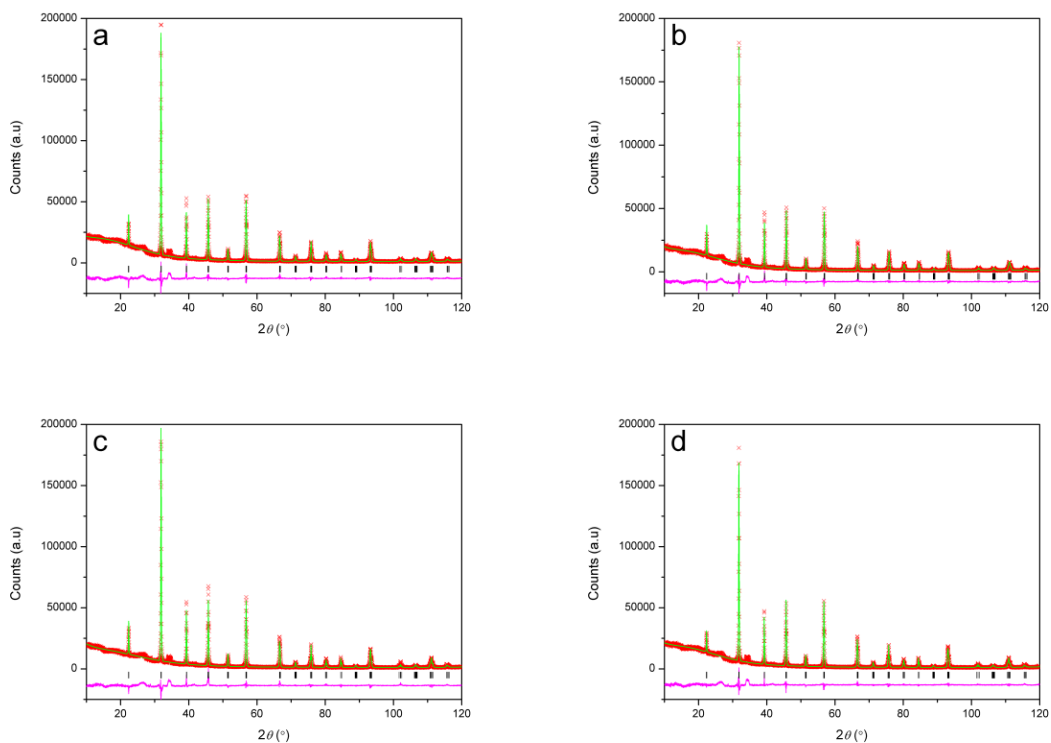

Fig S6, Fitted X-ray powder diffraction profiles for BST samples measured at -25 °C, with grain sizes of (a) 0.8 μm, (b) 1.2 μm, (c) 10 μm and (d) 25 μm. Observed (red crosses), calculated (green solid line) and difference (purple solid line) profiles are shown, along with the Bragg peak positions (black markers). Note additional peaks at approximately 26° and 35° 2θ are due to the sample holder.

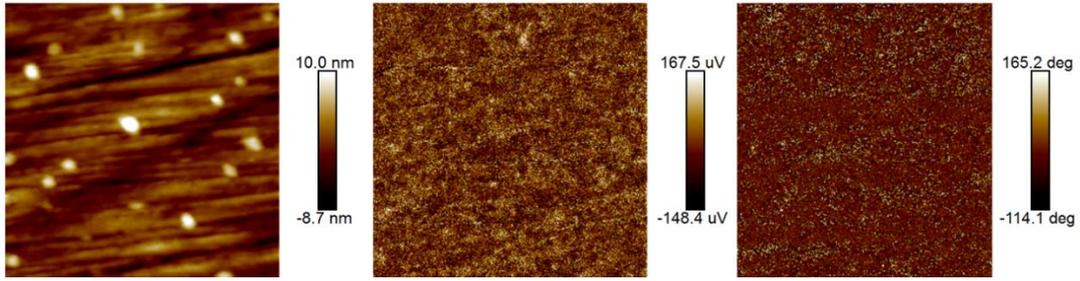

Fig. S7. PFM images of a BST sample with 10 μm grain size, (a) topography image, (b) amplitude image and (c) phase image.

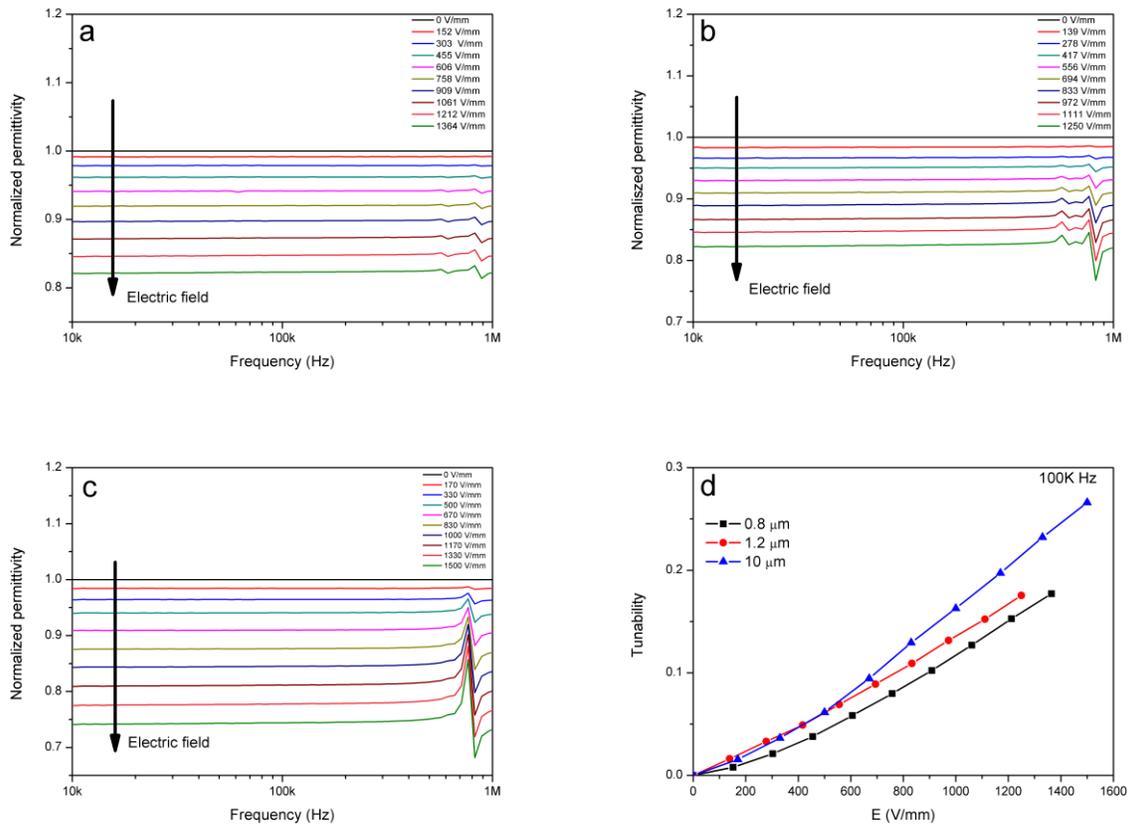

Fig. S8. (a-c) Variation of normalised dielectric permittivity under electric field for BST compositions with grain sizes of (a) 0.8 μm, (b) 1.2 μm and (c) 10 μm; (d) tunability variation with electric field at 100 kHz.

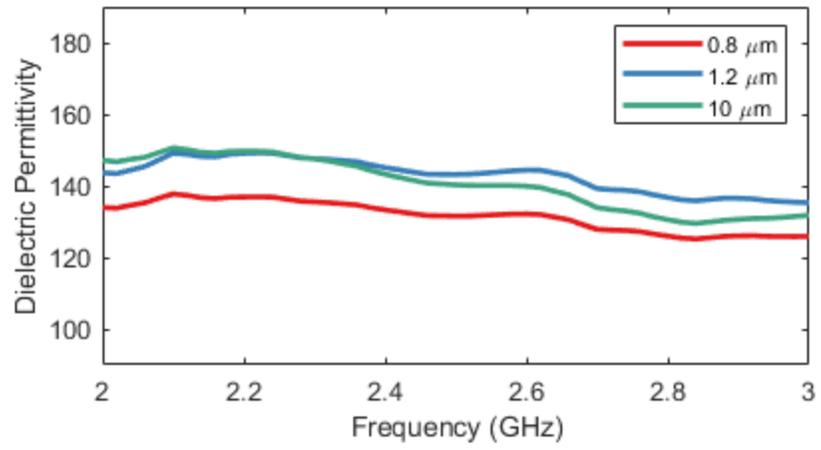

Fig. S9. The dielectric permittivity of the BST ceramic samples, as measured at microwave frequencies.